# Importance of the V 3d – O 2p hybridization in the Mott-Hubbard material $V_2O_3$

R.J.O. Mossanek and M. Abbate*

*Departamento de Física, Universidade Federal do Paraná,*
*Caixa Postal 19091, 81531-990 Curitiba PR, Brazil*

We studied the changes in the electronic structure of $V_2O_3$ using a cluster model. The calculations included fluctuations from the coherent band in the metallic phase, and non-local Mott-Hubbard fluctuations in the insulating phase. The incoherent structure is mostly related to the usual ligand screening channel ($3d^2\underline{L}$). The coherent peak in the metallic phase corresponds to coherent band fluctuations ($3d^2\underline{C}$). The non-local screened state in the insulating phase ($3d^2\underline{D}$) appears at higher energies, opening the band gap. The photon energy dependence of the spectra is mostly due to the relative V 3d and O 2p cross sections. The present model reproduces also the observed changes in the V 1s core-level spectra. The above results suggest that the Mott-Hubbard transition in $V_2O_3$ requires a multi-band model.

Many early transition metal oxides present a distinct metal-insulator transition (MIT) [1]. The particularly interesting MIT in $V_2O_3$ can be triggered by temperature, pressure, or doping [2]. Pure $V_2O_3$ goes from an antiferromagnetic insulator phase below 150-170 K, to a paramagnetic metal phase at higher temperatures [3-4]. The physics of this material is usually considered the paradigm of the Mott-Hubbard transition.

The first attempt to explain the MIT in $V_2O_3$ involved a mixed spin-orbital model [5]. The optical conductivity was investigated using the DMFT approach [6]. The electronic structure of $V_2O_3$ was calculated using the LDA+U and DMFT methods [7,8]. The orbital ordering was studied using RXS [9,10], neutron [11], and XAS [12]; these experimental results were explained in terms of spin coupling within the V-V pairs [13].

The electronic structure of $V_2O_3$ was studied using PES [14,15] and XAS [16]. More recent attempts involved combined PES-XAS [17] and XPS-XAS [18] studies. The photoemission spectra of metallic $V_2O$ shows a prominent quasi-particle peak [19]; the large intensity of this peak was attributed to larger bulk sensitivity at high photon energies. The satellite in the core level spectra was attributed to screening from the coherent band [20].

In this work, we studied the changes in the electronic structure of $V_2O_3$ using a cluster model. The calculation includes the screening from the coherent band in the metallic phase [20], and a distinct Mott-Hubbard screening in the insulating phase [21]. A similar method was recently used to study the electronic structure of metallic and insulating $VO_2$ [22]. The calculation reproduces the changes in the valence band and core level spectra with temperature, as well as the changes in the valence band spectra as a function of photon energy.

The above results suggest that the interpretation of the PES spectra should be revised. The incoherent structure is usually attributed to the remnant of the lower Hubbard band [6,8]. The present calculation shows that it corresponds mostly to a ligand screening state. The photon energy dependence is usually ascribed to the relative surface to bulk contribution [19]. The results here suggest that it is mainly due to the relative cross section of the V 3d – O 2p states.

Further, the standard DMFT approach does not include the O 2p states explicitly, and thus cannot explain the observed charge transfer satellites in the core level spectra. Whereas the present method explain not only the charge transfer satellites, but also the lower binding energy satellites due to the screening from the coherent band. This indicates that the Mott-Hubbard transition in $V_2O_3$ requires a V 3d – O 2p multi-band model.

The cluster considered here was composed by a $V^{3+}$ ion surrounded by an $O^{2-}$ octahedron. The cluster model was solved using the configuration interaction method. The main parameters of the model are the charge-transfer energy $\Delta$, the Mott-Hubbard energy U, the core-hole potential Q (Q = U/0.83), and the p-d transfer integral $T_\sigma$ [23]. The multiplet splitting was given in terms of the crystal field parameter 10Dq, and the p-p transfer integral $pp\pi$-$pp\sigma$.

The calculation includes the screening from the coherent band in the metallic phase [20]. The screening charge, in this case, comes from a delocalized state at the Fermi level. The metallic ground state was expanded in the $3d^2$, $3d^3\underline{C}$, $3d^3\underline{L}$, $3d^4\underline{C}^2$, $3d^4\underline{CL}$, $3d^4\underline{L}^2$, etc. configurations, where $\underline{C}$ denotes a hole in the coherent band, and $\underline{L}$ designates a hole in the O 2p band. The additional parameters are the coherent charge-transfer $\Delta^*$ and the transfer integral $T^*$ [20]. The value of $\Delta^*$ is roughly half the width of the occupied V 3d band, whereas $T^*$ is related to second order inter-cluster hopping processes. The configurations and energy levels of the metallic ground state are listed in Table I.

The calculation in the insulating phase included Mott-Hubbard charge fluctuations. The screening charge, in this phase, comes from a single nearest-neighbor $V^{3+}$ ion [21]. The insulating ground state was expanded in the $3d^2$, $3d^3\underline{D}$, $3d^3\underline{L}$, $3d^4\underline{D}^2$, $3d^4\underline{DL}$, $3d^4\underline{L}^2$, etc. configurations, were $\underline{D}$ denotes a hole in the neighboring $V^{3+}$ site. The Mott-Hubbard fluctuation energy becomes U, and the inter-cluster transfer T* was kept the same (because the lattice distortion in the insulating phase is small not affecting inter-cluster hopping). The configurations and energy levels of the insulating ground state are listed in Table II.

The removal (core-level) state was obtained by removing a valence (core-level) electron from the ground state. Finally, the corresponding spectral weight was calculated using the sudden approximation [23].

The model parameters used in the present calculation were: $\Delta = 5.0$ eV, $U = 3.5$ eV, $Q = 4.2$ eV, $T_\sigma = 1.75$ eV, $10Dq = 1.8$ eV, and $pp\pi$-$pp\sigma = 1.0$ eV. These values gave the best agreement with the experiment, and are also in excellent agreement with previous estimates [24]. The additional parameters were: $\Delta^* = 0.90$ eV ($U = 3.5$ eV), and $T^* = 0.20$ eV (0.20 eV) for the metallic (insulating) phase. The transfer integrals $T_\sigma$ and T* were increased by 10% in the core-level calculations.

The metallic $V_2O_3$ ground state is formed mainly by the $3d^2$ (52%), $3d^3\underline{L}$ (38%), and $3d^3\underline{C}$ (2.5%) configurations. The insulating $V_2O_3$ ground state is formed mostly by the $3d^2$ (53%), $3d^3\underline{L}$ (40%), and $3d^3\underline{D}$ (1.5%) configurations. The local screening configuration ($3d^3\underline{L}$) dominates the ground state, whereas the non-local screening configurations ($3d^3\underline{C}$ and $3d^3\underline{D}$) are much weaker. This is attributed to the relatively larger value of the local $T_\sigma$ compared to the non-local T* transfer.

Figure 1 shows the V 3d removal spectra of metallic and insulating $V_2O_3$; the corresponding spectra were decomposed in the main final configurations. In the metallic phase, the coherent peak, about –0.3 eV, is mostly formed by the $3d^2\underline{C}$ configuration (40%), whereas the incoherent part, around –1.2 eV, is mainly formed by the $3d^2\underline{L}$ configuration (37%). There is also a lower energy state, around –7.4 eV, which is mostly formed by the $3d^1$ configuration (56%). The existence of this state is reflected in the resonant photoemission spectra of $V_2O_3$ [14,15].

The results show that both the coherent and incoherent structures are *well-screened* states. The screening charge in the coherent part comes from the coherent band ($3d^2\underline{C}$), whereas the screening fluctuations in the incoherent part come from the ligand ($3d^2\underline{L}$). On the other hand, the state around –7.4 eV corresponds to the *poorly-screened* state ($3d^1$). The lowest energy fluctuation, in this phase, is from a $3d^2\underline{C}$ to a $3d^3$ configuration (d-d type). The material can be classified as a strongly V 3d – O 2p hybridized Mott-Hubbard system.

In the insulating phase, the non-local screened configuration becomes $3d^2\underline{D}$ instead of $3d^2\underline{C}$. The coherent part around –0.3 eV (~ –$\Delta$*), dominated mainly by the $3d^2\underline{C}$ configuration (40%), is replaced by a peak about –3.3 eV (~ –U), given mostly by the $3d^2\underline{D}$ configuration (53%). The incoherent structure, around –1.1 eV, is still mostly formed by the $3d^2\underline{L}$ configuration (47%). Finally, the low energy state about –7.6 eV is mostly formed by the $3d^1$ state (58%).

The insulating state is due to the transfer of spectral weight, from the coherent part ($3d^2\underline{C}$), around –$\Delta$*, to the non-local screened peak ($3d^2\underline{D}$), about –U. The band gap increases with the value of U as it should be in the Mott-Hubbard regime. However, the lowest energy fluctuation is now from a $3d^2\underline{L}$ to a $3d^3$ configuration (p-d type), which characterizes a strongly V 3d – O 2p hybridized charge-transfer regime. This illustrates that the classification of these kind of materials is not trivial, and further that the actual regime may even change across a metal-insulator transition.

The coherent structure, mostly $3d^2\underline{C}$, has an almost pure V 3d character, whereas the incoherent feature, mainly $3d^2\underline{L}$, is of mixed O 2p – V 3d character. The $3d^2\underline{C}$ final state configuration can only be obtained by removing a V 3d electron from the $3d^3\underline{C}$ ground state configuration. But the $3d^2\underline{L}$ final state configuration can be reached by removing either a V 3d electron (from $3d^3\underline{L}$) or an O 2p electron (from $3d^2$).

Figure 2 shows the relative weight of the V 3d and O 2p contributions to the removal spectra, compared to photoemission spectra taken from Ref. 17. The discrete states were convoluted with a Gaussian to simulate the resolution and band dispersion. The relative cross sections of the V 3d and O 2p levels were adjusted to a photon energy of 50 eV [25]. The calculation results are in excellent agreement with the experimental spectra, and show that the incoherent structure contains considerable O 2p character.

A recent high energy photoemission study of $V_2O_3$ showed a prominent coherent feature [19]. The enhancement of the coherent part was attributed to its relatively larger bulk character, whereas the larger contribution to the incoherent structure would come from the surface. Figure 3 shows the calculated photoemission spectra at 60 eV and 700 eV, taking into account the relative V 3d and O 2p cross sections [25]. The calculated results are in reasonably good agreement with the experimental spectra taken from Ref. 19. The changes can be thus mostly attributed to the relative energy dependence of the cross section. The relatively small discrepancy in the comparison is attributed to a residual surface vs. bulk effect.

Figure 4 shows the calculated core-level spectra of $V_2O_3$ in both phases, compared to the experimental V 1s photoelectron spectra taken from Ref. 20. The main structure is dominated by the *well-screened* $\underline{c}3d^3\underline{L}$ configuration, whereas the satellite peak is mostly formed by the *poorly-screened* $\underline{c}3d^2$ configuration. The low energy shoulder in the metallic phase is mainly given by the non-local $\underline{c}3d^3\underline{C}$ configuration, and appears at lower energies because $\Delta^* < \Delta$. The non-local screened state in the insulating phase is formed mainly by the $\underline{c}3d^3\underline{D}$ configuration, and appears at higher energies because $U > \Delta$.

The calculated removal and core-level spectra present the same qualitative trends. Namely, they are formed by low energy *well-screened* states ($3d^2\underline{L}$ or $\underline{c}3d^3\underline{L}$), and high energy *poorly-screened* states ($3d^1$ or $\underline{c}3d^2$). The lowest energy state in the metallic phase is related to coherent fluctuations ($3d^2\underline{C}$ or $\underline{c}3d^3\underline{C}$), which are replaced in the insulating phase by non-local screened states ($3d^2\underline{D}$ or $\underline{c}3d^3\underline{D}$). The present model provides a consistent description of both the valence and core-level spectra, whereas a single-band model would be unable to explain the satellites in the core-level spectra.

In conclusion, we studied the changes in the electronic structure of $V_2O_3$ using a cluster model. The main differences can be related to the different non-local screening channels in each phase. In the metallic phase, the non-local coherent state ($3d^2\underline{C}$) appears close to the Fermi level, because $\Delta^*$ is relatively small. In the insulating phase, the non-local Mott-Hubbard state ($3d^2\underline{D}$) appears at higher energies, because U is relatively large. The transfer of spectral weight from the $3d^2\underline{C}$ and $3d^2\underline{D}$ states produces the opening of the band gap. The photon energy dependence of the spectra is mostly due to the relative V 3d – O 2p cross sections. Finally, the present model reproduces the changes in both the valence band and core-level spectra.

| Metallic Ground States | |
|---|---|
| Configuration | Energy Level |
| $3d^2$ | 0 |
| $3d^3\underline{C}$ | $\Delta^*$ |
| $3d^3\underline{L}$ | $\Delta$ |
| $3d^4\underline{C}^2$ | $2\Delta^* + U$ |
| $3d^4\underline{CL}$ | $\Delta^* + \Delta + U$ |
| $3d^4\underline{L}^2$ | $2\Delta + U$ |

**Table I:** Configurations and energy levels of the metallic ground state.

| Insulating Ground States | |
|---|---|
| Configuration | Energy Level |
| $3d^2$ | 0 |
| $3d^3\underline{D}$ | U |
| $3d^3\underline{L}$ | $\Delta$ |
| $3d^4\underline{D}^2$ | 3U |
| $3d^4\underline{DL}$ | $\Delta + 2U$ |
| $3d^4\underline{L}^2$ | $2\Delta + U$ |

**Table II:** Configurations and energy levels of the insulating ground state.

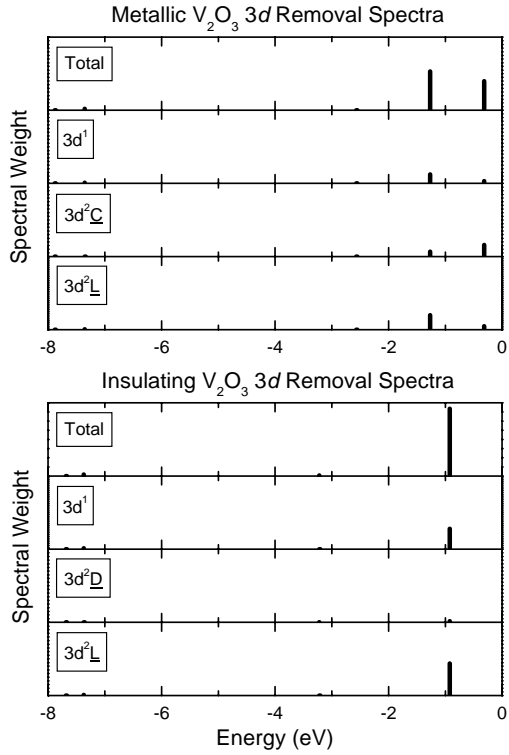

**Figure 1:** V *3d* removal spectra of metallic and insulating $V_2O_3$ projected on the main final state configurations.

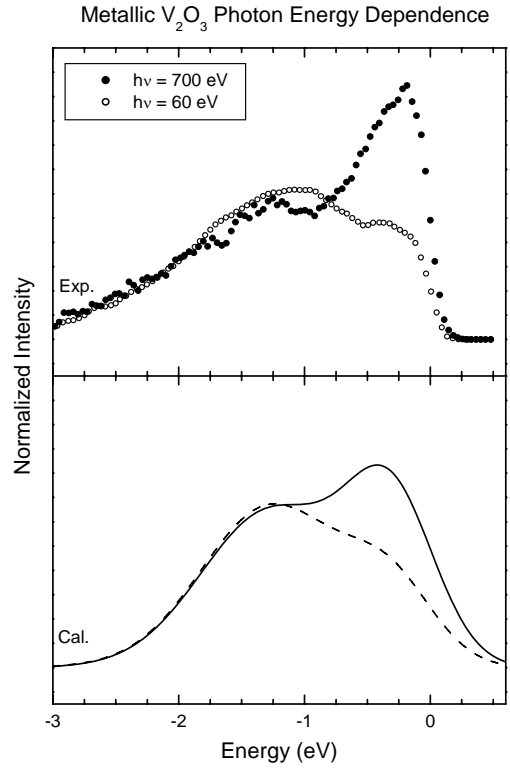

**Figure 3:** Calculated photon energy dependence compared to experimental spectra taken from Ref. 19.

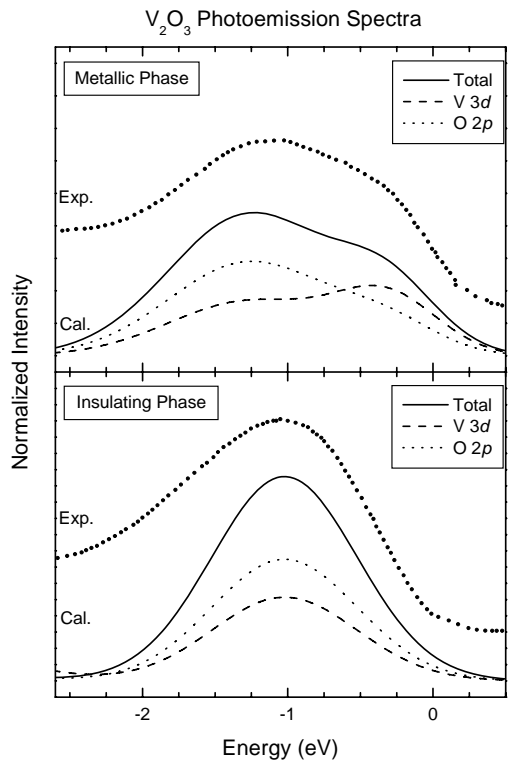

**Figure 2:** Calculated removal spectra with V *3d* and O *2p* contributions compared to photoemission spectra taken from Ref. 17.

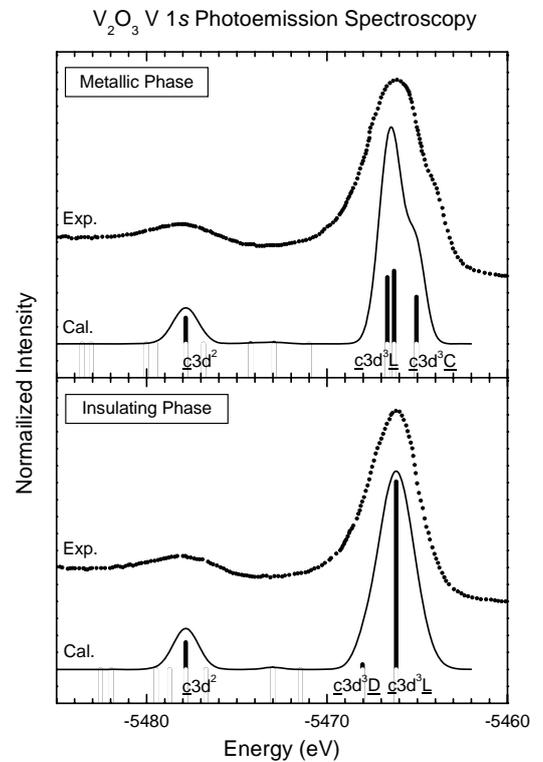

**Figure 4:** Calculated V *1s* core-level compared to experimental spectra taken from Ref. 20.